\documentclass[a4paper,12pt]{article}
\usepackage{amssymb,float,amsmath,amsfonts,amsthm,cases,cite}
\usepackage[normalem]{ulem}
\usepackage[pdfencoding=auto]{hyperref}
\usepackage{xcolor}
\usepackage{hyperref}
\definecolor{linkcolor}{HTML}{799B03}
\definecolor{urlcolor}{HTML}{799B03}
\hypersetup{pdfstartview=FitH,linkcolor=linkcolor,urlcolor=urlcolor,colorlinks=true}
\usepackage{graphicx}
\def\[{\begin{equation}}
\def\]{\end{equation}}

\makeatletter
\newcommand*{\myfnsymbolsingle}[1]{%
  \ensuremath{%
    \ifcase#1
    \or 
      *%
    \or 
      \dagger
    \or 
      \ddagger
    \or 
      1
    \or 
      2
    \or   
      3
    \or
      4
    \or 
      5
    \or
      6
    \or   
      7
    \or
      8
    \or 
      9
    \or
      10
    \or   
      11
    \or
      12
    \or 
      13
    \or
      14
    \else 
      \@ctrerr  
    \fi
  }%
}   
\makeatother

\usepackage{alphalph}
\newalphalph{\myfnsymbolmult}[mult]{\myfnsymbolsingle}{}

\renewcommand*{\thefootnote}{%
  \myfnsymbolmult{\value{footnote}}%
}

\begin{document}	
\vspace{10pt}

\begin{center}
{\LARGE \bf Bounce beyond Horndeski with GR asymptotics and $\gamma$-crossing.}

\vspace{20pt}

S. Mironov$^{a,c}$\footnote{sa.mironov\_1@physics.msu.ru},
V. Rubakov$^{a,b}$\footnote{rubakov@minus.inr.ac.ru}
V. Volkova$^{a,b}$\footnote{volkova.viktoriya@physics.msu.ru}
\renewcommand*{\thefootnote}{\arabic{footnote}}
\vspace{15pt}

$^a$\textit{Institute for Nuclear Research of the Russian Academy of Sciences,\\
60th October Anniversary Prospect, 7a, 117312 Moscow, Russia}\\
\vspace{5pt}

$^b$\textit{Department of Particle Physics and Cosmology, Physics Faculty,\\
M.V. Lomonosov Moscow State University,\\
Vorobjevy Gory, 119991 Moscow, Russia}

$^c$\textit{Institute for Theoretical and Experimental Physics,\\
Bolshaya Cheriomyshkinskaya, 25, 117218 Moscow, Russia}
\end{center}

\vspace{5pt}

\begin{abstract}
It is known that beyond Horndeski theory admits healthy bouncing 
cosmological solutions. However, the constructions proposed so far 
do not reduce to General Relativity (GR) in either infinite past 
or infinite future or both. 
The obstacle is so called $\gamma$-crossing, which
off hand appears pathological. By working in the unitary gauge, 
we confirm
the recent observation by Ijjas~\cite{Ijjas:2017pei} that 
$\gamma$-crossing is,
in fact, healthy. On this basis we construct a spatially flat, 
stable bouncing Universe solution whose
asymptotic past and future are described by GR with
conventional massless scalar field.  
\end{abstract}

\section{Introduction and summary}
Horndeski theories~\cite{Horndeski:1974wa, Fairlie:1991qe, Nicolis:2008in, Deffayet:2009wt} are the most general 
scalar--tensor theories whose equations of motion are second order 
despite the presence of higher derivatives in the Lagrangian. 
There is an extension of the general Horndeski theory, which is 
referred to as "beyond Horndeski"~\cite{Zumalacarregui:2013pma, Gleyzes:2014dya}. The difference between Horndeski 
and beyond Horndeski theories is that in the latter, the equations 
of motion are third order but still with no Ostrogradsky 
instabilities arising.
Further generalization is dubbed "DHOST" theories~\cite{Langlois:2015cwa, BenAchour:2016fzp, Langlois:2017mxy}. 

Horndeski and beyond Horndeski theories are widely used for 
constructing various cosmological solutions like cosmological bounce, 
Genesis, etc.~\cite{Creminelli:2006xe, Kobayashi:2010cm, Kobayashi:2011nu,Easson:2011zy, Cai:2012va, Koehn:2013upa,Pirtskhalava:2014esa,Qiu:2015nha,Kobayashi:2015gga,Wan:2015hya,Koehn:2015vvy,Ijjas:2016tpn}. Indeed, (beyond) Horndeski theories
are capable of violating the Null Energy Condition without obvious
pathologies (for a review see, e.g., Ref.~\cite{Rubakov:2014jja}).
In this paper we concentrate on the classical bouncing solutions. 
Even though numerous examples of spatially flat bouncing solutions were suggested
within the general Horndeski theory, it was shown in Refs.~\cite{Libanov:2016kfc,Kolevatov:2016ppi,Kobayashi:2016xpl} that 
bouncing solutions in this class of theories are plagued with gradient
instabilities, which arise if one considers the entire evolution. 
Attempts to evade this no-go theorem within the general
Horndeski theory result in 
either singularity or potential strong coupling.
It was found, however, that going beyond Horndeski enables one to satisfy 
the stability conditions and obtain a complete, healthy, 
spatially flat bouncing 
solutions~\cite{Cai:2016thi,  Creminelli:2016zwa , Kolevatov:2017voe, Cai:2017dyi   }. The only drawback left is that the 
suggested solutions do not have simple asymptotics. 
Namely, the bouncing solutions designed so far
do not reduce to General Relativity (GR) in either the asymptotic past or 
future or both. This feature appears dissatisfying.

As discussed in detail in Ref.~\cite{Kolevatov:2017voe}, 
the obstacle to having GR in both asymptotic past and future is the 
so called $\gamma$-crossing. The latter phenomenon has to do with the 
quadratic action for scalar perturbations in the unitary gauge. 
The coefficients there involve the denominator\footnote{This denominator is denoted by $\Theta$ in Refs.~\cite{Kobayashi:2016xpl, Kolevatov:2017voe}, 
$\gamma$ in Refs.~\cite{Ijjas:2016tpn, Ijjas:2017pei, Dobre:2017pnt}, and $A_4$ in the current work.},
and $\gamma$-crossing occurs when this denominator vanishes.
Barring fine-tuning, the coefficients in the unitary gauge quadratic
action diverge at $\gamma$-crossing
\footnote{It has been noted in Ref.~\cite{Dobre:2017pnt} that zero denominator 
corresponds to the interchange of the solution branches 
for the Hubble parameter in the Friedmann equation.}.
Until recently, this has been considered unacceptable, and bouncing 
solutions obtained so far avoided $\gamma$-crossing. The price to pay
was the strong modification of gravity at early and/or 
late times.

The issue of $\gamma$-crossing has been recently discussed in Ref.~\cite{Ijjas:2017pei} 
from a new perspective. It has been shown that by choosing the 
Newtonian gauge one obtains the linearized equations for metric perturbations
{\itshape without} any denominator. Hence, there is no problem 
with $\gamma$-crossing in the Newtonian gauge at all. 
This apparent discrepancy between the unitary and Newtonian gauges is puzzling.
On the one hand, one might suspect that the stability analysis in terms of the unitary gauge 
set of variables cannot be carried out around $\gamma$-crossing, 
since the linearized equations become singular at this point. 
On the other hand, in gauge invariant theories like beyond Horndeski,
gauge fixing should not affect the physics. 

In this paper we show explicitly that despite the seeming problem with 
$\gamma$-crossing in the unitary gauge, the solutions for all the 
perturbation variables in this gauge are regular for any 
value of the denominator including zero. In other words, the singularities
are present only in the linearized equations but not in their solutions. 
Hence, we safely allow for $\gamma$-crossing. This is in accordance with
the general conclusion by Ijjas~\cite{Ijjas:2017pei} , even though
we disagree with her claim that the unitary gauge is ill-defined 
at $\gamma$-crossing. 
As we further discuss in this paper, even though $\gamma$-crossing
is a healthy phenomenon, it does not enable one to circumvent 
the above no-go theorem in the Horndeski theory.

Most importantly, healthy $\gamma$-crossing enables us to construct a 
complete, stable, spatially flat bouncing solution in 
beyond Horndeski theory
%
%
whose past {\itshape and} future asymptotics are described by a theory 
of a conventional massless scalar field and GR. We give an explicit example 
of such a bouncing solution and check its stability 
during entire evolution.

This paper is organized as follows. We introduce the beyond Horndeski 
theory and give the basic formulas of the linearized perturbation 
theory in Sec.~\ref{Basics}. In Sec.~\ref{Solution} we demonstrate 
that the solutions for linearized perturbations in the unitary gauge 
are indeed non-singular for any value of the denominator including zero,
i.e., that $\gamma$-crossing is not pathological. 
We also revisit the no-go argument for the general Horndeski theory and 
stress that healthy $\gamma$-crossing does not help to evade the no-go theorem.
In Sec.~\ref{Bounce} we present an explicit example of the 
healthy bouncing solution in beyond Horndeski theory,
which connects two asymptotics with a massless scalar field and 
the conventional Einstein gravity.

\section{Perturbations in Horndeski theory and beyond}
\label{Basics}

In what follows we consider both the general Horndeski and 
beyond Horndeski cases. The Lagrangian of the beyond Horndeski
theory has the form (mostly negative metric signature):
\begin{subequations}
	\label{lagrangian}
	\begin{align}
	&S=\int\mathrm{d}^4x\sqrt{-g}\left(\mathcal{L}_2 + \mathcal{L}_3 + \mathcal{L}_4 + \mathcal{L}_5 + \mathcal{L_{BH}}\right),\\
	&\mathcal{L}_2=F(\pi,X),\\
	&\mathcal{L}_3=K(\pi,X)\Box\pi,\\
	&\mathcal{L}_4=-G_4(\pi,X)R+2G_{4X}(\pi,X)\left[\left(\Box\pi\right)^2-\pi_{;\mu\nu}\pi^{;\mu\nu}\right],\\
	&\mathcal{L}_5=G_5(\pi,X)G^{\mu\nu}\pi_{;\mu\nu}+\frac{1}{3}G_{5X}\left[\left(\Box\pi\right)^3-3\Box\pi\pi_{;\mu\nu}\pi^{;\mu\nu}+2\pi_{;\mu\nu}\pi^{;\mu\rho}\pi_{;\rho}^{\;\;\nu}\right],\\
	&\mathcal{L_{BH}}=F_4(\pi,X)\epsilon^{\mu\nu\rho}_{\quad\;\sigma}\epsilon^{\mu'\nu'\rho'\sigma}\pi_{,\mu}\pi_{,\mu'}\pi_{;\nu\nu'}\pi_{;\rho\rho'}+
	\\\nonumber&\qquad+F_5(\pi,X)\epsilon^{\mu\nu\rho\sigma}\epsilon^{\mu'\nu'\rho'\sigma'}\pi_{,\mu}\pi_{,\mu'}\pi_{;\nu\nu'}\pi_{;\rho\rho'}\pi_{;\sigma\sigma'},
	\end{align}
\end{subequations} 
where $\pi$ is the scalar field (sometimes dubbed generalized Galileon),
$X=g^{\mu\nu}\pi_{,\mu}\pi_{,\nu}$, $\pi_{,\mu}=\partial_\mu\pi$, 
$\pi_{;\mu\nu}=\triangledown_\nu\triangledown_\mu\pi$, 
$\Box\pi = g^{\mu\nu}\triangledown_\nu\triangledown_\mu\pi$, 
$G_{4X}=\partial G_4/\partial X$, etc. 
The Horndeski theory corresponds to $F_4(\pi,X)=F_5(\pi,X)=0$.

In this and next sections we concentrate mostly on the scalar sector of 
perturbations about the spatially flat FLRW background.
In this case the ADM decomposition of the linearized metric 
has the following form:
\[
\label{perturbations}
ds^2 = (1+2\alpha)dt^2 - \partial_i\beta \;dt dx^i - a^2 (1 + 2\zeta \delta_{ij} + 2 \partial_i\partial_j E)dx^i dx^j.
\]    
The scalar field perturbation is denoted by $\delta\pi=\chi$. 
Without loss of generality we partly use the gauge freedom and gauge away
the longitudinal component $\partial_i\partial_j E$ from the very beginning. 
Then, the quadratic action for the scalar perturbations has the form
\[
\label{full_quadr_action}
\begin{aligned}
S^{(2)} = \int &\mathrm{d}t\,\mathrm{d}^3x\,a^3 \Bigg (
A_1\:\dot{\zeta}^2 + A_2 \:\:\dfrac{(\overrightarrow{\nabla}\zeta)^2}{a^2} + A_3\: \alpha^2 + A_4\: \alpha\dfrac{\overrightarrow{\nabla}^2\beta}{a^2} + A_5\: \dot{\zeta}\dfrac{\overrightarrow{\nabla}^2\beta}{a^2} + A_6\: \alpha\dot{\zeta} \\
+A_7&\: \alpha\:\dfrac{\overrightarrow{\nabla^2}\zeta}{a^2}
+ A_8\:\alpha\dfrac{\overrightarrow{\nabla}^2\chi}{a^2} + A_9\: \dot{\chi}\dfrac{\overrightarrow{\nabla}^2\beta}{a^2} + A_{10}\:\chi\ddot{\zeta} + A_{11}\:\alpha\dot{\chi} 
+A_{12}\:\chi\dfrac{\overrightarrow{\nabla}^2\beta}{a^2}
+ A_{13}\:\chi\dfrac{\overrightarrow{\nabla}^2\zeta}{a^2} \\
+A_{14}&\:\dot{\chi}^2 + A_{15}\:\dfrac{(\overrightarrow{\nabla}\chi)^2}{a^2}  + B_{16}\:\dot{\chi}\dfrac{\overrightarrow{\nabla}^2\zeta}{a^2}
+A_{17}\: \alpha\chi + A_{18}\:\dot{\zeta}\chi + A_{19}\:\zeta \chi +A_{20}\:\chi^2 \Bigg),
\end{aligned}
\]
where an overdot stands for derivative with respect to cosmic time $t$, 
and coefficients $A_i$ are expressed in terms of the Lagrangian functions 
and their derivatives. Their explicit expressions are collected in Appendix A. 
Note that the terms $\alpha\zeta$ and $\zeta^2$ have vanishing 
coefficients thanks to the background equations. The correspondence 
between our coefficients $A_i$ and those in Refs.~\cite{Kobayashi:2016xpl, Kolevatov:2017voe}  is
\[
\label{dictionary}
A_1 = -3 \mathcal{\hat{G}_T}, \;\; A_2 = \mathcal{F_T}, \;\; A_3 = \Sigma, \;\;
A_4  =-2 \Theta, \;\; A_5 = 2 \mathcal{\hat{G}_T}, \;\; A_6 = 6 \Theta, \;\; 
A_7 = - 2 \mathcal{G_T}.
\]
Also, the coefficient $A_4$ is denoted in Refs.~\cite{Ijjas:2017pei, Ijjas:2016tpn ,Dobre:2017pnt} by
\[
\label{gamma}
A_4 = 2 \gamma.
\]
It is important for what follows that the coefficients $A_5$ 
and $(-A_7)$ differ only by beyond Horndeski terms. Explicitly,
see Appendix A,
\[
\label{A5A7}
A_5 + A_7 = -B_{16}\dot{\pi} = 4F_4 \dot{\pi}^4 + 12 H F_5 \dot{\pi}^5.
\]
The quadratic action \eqref{full_quadr_action} is invariant under 
the residual gauge transformations:
\[
\label{gauge}
\alpha \to \alpha + \dot{\xi_0},\quad \beta \to \beta - \xi_0, \quad \chi \to
\chi + \xi_0\dot{\pi},\quad \zeta \to \zeta + \xi_0 H,
\]
where $H$ is the Hubble parameter and $\xi_0$ is the gauge function.

Lapse ($\alpha$) and shift ($\beta$) variables are non-dynamical, 
and variation of the action~\eqref{full_quadr_action} with respect 
to them leads to the following constraints: 
\begin{subequations}
\label{constraints}
\begin{align}
\alpha &= -\dfrac{1}{A_4}\left(A_5\;\dot{\zeta}+A_{9}\;\dot{\chi}+A_{12}\;\chi\right), \\
\dfrac{\overrightarrow{\nabla}^2\beta}{a^2} &= -\dfrac{1}{A_4}\left(A_7\; \dfrac{\overrightarrow{\nabla}^2\zeta}{a^2} + A_8 \dfrac{\overrightarrow{\nabla}^2\chi}{a^2} + A_6\;\dot{\zeta} +A_{11}\;\dot{\chi} +A_{17}\; \chi\right) \\
&\nonumber+ \frac{2 A_3}{A_4^2}\left(A_5\;\dot{\zeta}+A_{9}\;\dot{\chi}+A_{12}\;\chi\right).
\end{align}
\end{subequations}
By utilizing the constraints~\eqref{constraints}, we integrate out 
$\alpha$ and $\beta$. We write the resulting action in terms of gauge
invariant combination of the curvature and scalar field perturbations:
\[
\label{inv_quadr_action}
\begin{aligned}
S^{(2)} = \int &\mathrm{d}t\,\mathrm{d}^3x\,a^3 \Bigg(
\mathcal{A} \cdot \left(\frac{d}{dt}\left[\zeta \dot{\pi} - \chi H\right]\right)^2 +
\mathcal{B} \cdot \left(\zeta \dot{\pi} - \chi H \right)^2 -
\mathcal{C} \left(\dfrac{\overrightarrow{\nabla}\zeta}{a} \dot{\pi} - \dfrac{\overrightarrow{\nabla}\chi}{a} H \right)^2
\Bigg)
\end{aligned},
\] 
where
\begin{subequations}
\label{mathcal_ABC}
\begin{align}
\label{mathcal_A}
\mathcal{A} &= \frac{1}{\dot{\pi}^2}\left( A_1 + \frac{A_3\cdot A_5^2}{A_4^2} - \frac{A_5\cdot A_6}{A_4}\right)\equiv \frac{1}{\dot{\pi}^2} \left( \frac{4}{9} \frac{A_1^2 \cdot A_3}{A_4^2} - A_1\right),\\
\label{mathcal_B}
\mathcal{B} &= \frac{\mathcal{A} \dddot{\pi} + \dot{\mathcal{A}}  \ddot{\pi} + 3 \mathcal{A} H \ddot{\pi}}{\dot{\pi}},\\
\label{mathcal_C}
\mathcal{C} &= \frac{1}{\dot{\pi}^2}\left(\dfrac{1}{a} \frac{d}{dt}\left[ \frac{a A_5 \cdot A_7}{2 A_4}\right] - A_2 \right).
\end{align}
\end{subequations}

The potentially problematic situation occurs if the coefficient $A_4$
crosses zero. Following Refs.~\cite{Ijjas:2017pei, Ijjas:2016tpn, Dobre:2017pnt} we call it 
$\gamma$-crossing, see eq.~\eqref{gamma}. 
Indeed, according to constraints~\eqref{constraints}, both 
$\alpha$ and $\beta$ appear singular when $A_4 = 0$. 
Moreover, the coefficients $\mathcal{A}$, $\mathcal{B}$ and 
$\mathcal{C}$ in the quadratic action~\eqref{inv_quadr_action} 
hit singularity, making the stability analysis tricky. 

However, it has been shown in Ref.~\cite{Ijjas:2017pei}, that in the 
Newtonian gauge, the solutions for the variables 
$\alpha$, $\zeta$ and $\chi$ are regular for all values 
of $A_4$, including zero. This implies that the 
solutions for all scalar perturbations are everywhere 
regular in any other gauge. To see explicitly that this
is indeed the case, we carry out in Sec.~\ref{Solution}
calculations analogous to Ref.~\cite{Ijjas:2017pei} but in the 
{\itshape unitary} gauge and show that the solutions for all 
variables in the unitary gauge, namely, $\zeta$, 
$\alpha$ and $\beta$, are in fact regular at $\gamma$-crossing.

\section{$\gamma$-crossing}
\label{Solution}
\subsection{Solutions for metric perturbations in the unitary gauge}
In this section we obtain the solutions for $\zeta$, $\alpha$ 
and $\beta$ in the unitary gauge and show that these are regular 
despite the seeming pathology of eqs.~\eqref{constraints} 
and action~\eqref{inv_quadr_action} at $\gamma$-crossing. 

As the first step, let us assume that $\alpha$ and $\beta$ are finite
and can be found from eqs.~\eqref{constraints} for any value 
of $A_4$ including zero (below we explicitly show that 
this assumption does hold). 
This enables one to legitimately obtain the quadratic 
action~\eqref{inv_quadr_action} in a standard manner. 
Upon imposing the unitary gauge $\chi =0$ in the action~\eqref{inv_quadr_action}, one obtains the
linearized equation for $\zeta$:
\begin{equation}
\label{zeta_lin_equation}
\mathcal{A} \dot{\pi}^2 \cdot \ddot{\zeta} + \left( \dot{\mathcal{A}}\dot{\pi}^2 + 2 \mathcal{A} \dot{\pi} \ddot{\pi} + 3 \mathcal{A} H \dot{\pi}^2 \right) \cdot \dot{\zeta} - \mathcal{C} \dot{\pi}^2 \cdot \frac{\overrightarrow{\nabla}^2\zeta}{a^2}= 0,
\end{equation}
In what follows, we keep track of the coefficient $A_4$ and its time
derivatives only. 
Making use of the definitions~\eqref{mathcal_ABC} and performing 
Fourier transformation, we write eq.~\eqref{zeta_lin_equation} in
the following form:
\[
\label{final_zeta_equation}
\left( 1 + c_1 \cdot A_4^2\right) \cdot \ddot{\zeta} + \left( c_2 + c_3 \cdot A_4^2 - 2 \cdot \frac{\dot{A_4}}{A_4}\right) \dot{\zeta} +
\frac{k^2}{a^2} \left(c_4 \cdot \dot{A_4} +c_5 \cdot A_4 + c_6 \cdot A_4^2 \right) \cdot \zeta = 0,
\]
where $c_i$ are combinations of the coefficients $A_i$, $i \neq 4$.
These combinations are non-singular at $\gamma$-crossing.
Since for homogeneous background the coefficients $A_i$ are
functions of time only, so are the coefficients $c_i$ 
in eq.~\eqref{final_zeta_equation}.

To study the behavior of metric perturbations at $\gamma$-crossing, 
we choose the origin of time in such a way 
that $\gamma$-crossing occurs at $t=0$ and write
\begin{equation}
\label{theta_choice}
A_4 = C \cdot t + \dots,
\end{equation}
where $C$ is a constant and dots denote terms of higher order in $t$.

Let us first obtain the solutions to eq.~\eqref{final_zeta_equation} 
to the leading order.
Keeping only the dominant terms in the vicinity of $t=0$, we find
that eq.~\eqref{final_zeta_equation} is reduced to
\[
\label{reduced_zeta_eq}
\ddot{\zeta} - \frac{2}{t}\dot{\zeta} = 0,
\]
and the two solutions to this equation are
\[
\label{reduced_zeta_solution_0}
\zeta = \lambda = \mbox{const},
\]
and
\[
\label{reduced_zeta_solution_3}
\zeta = \delta \cdot t^3,\;\; \delta = \mbox{const}.
\]
Importantly, the corrections to the solution~\eqref{reduced_zeta_solution_0}
start with $t^2$:
\begin{equation}
\label{zeta-solution}
\zeta = \lambda \left(1 + \frac{C}{2} \frac{k^2}{a^2} c_4 \cdot t^2 +  \dots \right),
\end{equation} 
where the coefficient $c_4$ is, explicitly,
\[
\label{c_4}
c_4 = \frac34  \cdot \frac{A_7}{A_1 \cdot A_3}.
\] 
Thus, even though the linearized equation for $\zeta$ is singular 
at $t=0$, the solutions~\eqref{reduced_zeta_solution_0} and~\eqref{zeta-solution} are regular.

Let us see that the lapse and shift are regular as well. 
Making use of the relations $A_5 = -\frac23 A_1$ and $A_6 = - 3 A_4$
(see eqs.~\eqref{A_5} and~\eqref{A_6} in Appendix A) we write eqs.~\eqref{constraints} in the unitary gauge:
\begin{subequations}
\label{constraints_unitary_gauge}
\begin{align}
\label{a_eq}
\alpha &= -\dfrac{A_5}{A_4}\cdot \dot{\zeta}, \\
\label{b_eq}
-\dfrac{k^2}{a^2}\beta &= \dfrac{A_7}{A_4} \dfrac{k^2}{a^2}\cdot \zeta  + \left(3- \frac43\frac{A_1 A_3}{A_4^2} \right) \cdot \dot{\zeta},
\end{align}
\end{subequations}
Because of eq.~\eqref{zeta-solution}, the shift perturbation~\eqref{a_eq}
is obviously regular. Off hand, the right hand side of eq.~\eqref{b_eq}
is of order $t^{-1}$. However, the terms of order $t^{-1}$ cancel out:
in view of~\eqref{theta_choice},~\eqref{zeta-solution} and~\eqref{c_4},
we have
\[
\label{cancel_in_beta}
\dfrac{A_7}{A_4} \dfrac{k^2}{a^2}\cdot \zeta - \left(3- \frac43\frac{A_1 A_3}{A_4^2}\right)  \cdot \dot{\zeta} = \frac{1}{A_4}\left[ A_7 \dfrac{k^2}{a^2}\cdot \lambda - \frac43\frac{A_1 A_3}{A_4}\left( \dfrac{k^2}{a^2} c_4 \cdot C t \right)\lambda + O(t) \right] = O(1).
\]
Therefore, the shift perturbation $\beta$ is regular as well.
Moreover, it is now evident that the unitary and Newtonian gauges 
are related by a non-singular transformation. Indeed, moving from 
the unitary to Newtonian gauge amounts to gauging away $\beta$ and 
introducing $\chi$ back. Since $\beta$ is regular, the
corresponding gauge function $\xi_0$ in~\eqref{gauge} 
is regular as well.

Hence, we have explicitly shown that there is nothing wrong 
with letting $A_4$ to cross zero. It is still possible to analyse 
the stability in terms of the unitary gauge set of variables around 
this point. 

As discussed in Refs.~\cite{Kolevatov:2017voe, 
Ijjas:2017pei} , $\gamma$-crossing
is essential for constructing spatially flat bouncing solutions, 
which connect two asymptotic states of the Universe, in which
the field $\pi$ is a conventional scalar field
and gravity is described by conventional GR.
Indeed, in that case $G_4 \rightarrow 1/2$, $K, G_5, F_4, F_5 \rightarrow 0$ as $|t| \rightarrow \infty$, and, according to the explicit expression~\eqref{A_4} given in Appendix A,
$A_4 \rightarrow - 2 H$. 
Bouncing solution implies that the Hubble parameter, 
and hence $(- A_4)$, are negative at early times and positive 
at late times, so $A_4$ crosses zero somewhere. 
Since in many previous works $\gamma$-crossing was believed 
to be troublesome, the scenario with restored Einstein
gravity long before and after the bounce was not considered.
In the Sec.~\ref{Solution} we construct a specific example of 
this type of bouncing solution. 

\subsection{No-go theorem and $\gamma$-crossing}
\label{no_go}
\begin{figure}[H]\begin{center}\hspace{-1cm}
{\includegraphics[width=0.7\linewidth]
{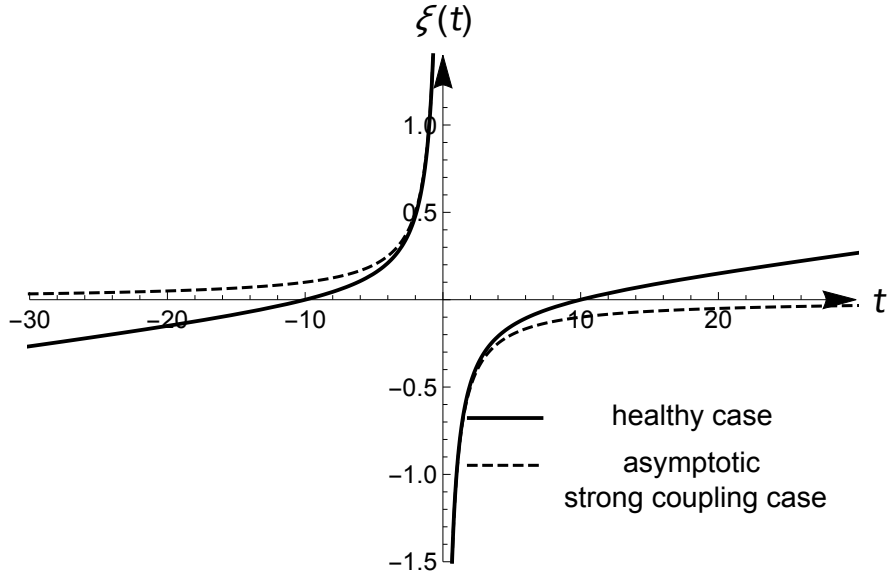}}\hspace{-1cm}
\caption{\footnotesize{$\xi(t)$ with $\gamma$-crossing (without loss of generality $\gamma$-crossing takes place at $t=0$). Dashed line
shows $\xi(t)$ which asymptotically tends to zero and 
faces strong
coupling problem
at both infinities: $\dot{\xi}\rightarrow 0$. Solid line represents 
a healthy behavior of $\xi$: $\dot{\xi}>\mbox{const}>0$ and $\xi$ 
crosses zero twice.}}
\label{asc}
\end{center}\end{figure}

Let us now briefly revisit the no-go theorem~\cite{Kobayashi:2016xpl} for the general 
Horndeski theory and emphasise that $\gamma$-crossing does not help
to evade this theorem. 
To this end, let us recall the form of the quadratic action
for tensor perturbations valid both in Horndeski and beyond 
Horndeski theories:
\[
\label{action_tensor}
\begin{aligned}
S^{(2)}_{tensor} = \int &\mathrm{d}t\,\mathrm{d}^3x\,a^3 \left[\dfrac{A_5}{2}\left(\dot{h}^T_{ik}\right)^2 - A_2 \dfrac{\left(\overrightarrow{\nabla} h_{ik}^T\right)^2}{a^2} 
\right],
\end{aligned}
\] 
where $h^T_{ik}$ denotes transverse traceless tensor perturbation.
The quadratic action in the scalar sector has the 
form~\eqref{inv_quadr_action}. To avoid ghost and gradient 
instabilities one requires $A_5 > 0$, $\mathcal{A} >0$ and $A_2 >0$, 
$\mathcal{C} >0$.  
The main no-go argument is based on the requirement of the absence 
of gradient instabilities in the scalar sector, i.e.
$\mathcal{C} > 0$.
Taking into account the positivity of both $A_2$ and $\mathcal{C}$ 
and using the definition of $\mathcal{C}$ in~\eqref{mathcal_C}, 
we write this requirement in the following form
\footnote{In addition to~\eqref{dictionary}, we note that our notations 
are related to those in Refs.~\cite{Kobayashi:2016xpl, Kolevatov:2017voe} as 
$\mathcal{A} = \mathcal{G_S}$, $\mathcal{C} = \mathcal{F_S}$.}: 
\[
\label{new_no-go}
\frac{d}{dt}\left[ \frac{a A_5 \cdot A_7}{2 A_4}\right] = a \cdot (\mathcal{C}\dot{\pi}^2 + A_2) > 0,
\]
The point is that 
\[
\label{ksi}
\xi = \frac{a A_5 \cdot A_7}{2 A_4}
\]
is, therefore,
a monotonously growing function.
If the theory at $t \to \pm \infty$
is more or less
{\itshape conventional}, the coefficients $A_2$ and  
$\mathcal{C}\dot{\pi}^2$
in \eqref{action_tensor} and 
\eqref{inv_quadr_action} tend to positive
constants as  $t \to \pm \infty$
(or, more generally, are bounded from below by positive constants);
then $\dot{\xi} > \mbox{const} > 0$ for all times and, thus, $\xi$
cannot asymptotically tend to a constant value or zero.
Hence, $\xi$  necessarily crosses zero somewhere during 
the evolution. Importantly, this occurs both in Horndeski and beyond
Horndeski theories and irrespectively
of the $\gamma$-crossing, at which $|\xi| = \infty$.
This situation is shown in Fig.~\ref{asc} by solid lines.

In the case of the general Horndeski theory, one has 
\[
\label{A7toA5}
A_5 = - A_7,
\]
see~\eqref{A5A7}, and $A_5$ has to be positive 
to avoid ghost instabilities in the tensor sector. So,
{\it in Horndeski theory with conventional asymptotics} $\xi$ necessarily crosses zero, and this occurs
when
$A_4 \rightarrow \infty$. This scenario
of infinite $A_4$ implies the singularity in the 
classical solution; we get back to the no-go theorem
of Refs.~\cite{Libanov:2016kfc,Kobayashi:2016xpl} which, we
emphasize, holds even in the presence of $\gamma$-crossing.

There are in principle two ways out in Horndeski theory.
Sticking to  conventional asymptotics, and hence $\xi$ crossing zero,
one can have $A_4 = 0$ {\it and}  $A_5 = - A_7=0$ at this
point~\cite{Kobayashi:2016xpl, Ijjas:2016tpn} (we give concrete example below in Fig.~\ref{finetune}).
This case is not only fine-tuned,
but also faces strong coupling problem in the tensor sector, 
see eq.~\eqref{action_tensor}. The second possibility
is to give up the conventional asymptotics of the action for quadratic
perturbations and consider models
with $A_2, \: \mathcal{C} \rightarrow 0$ as $t\to -\infty$
and/or $t\to +\infty$. This scenario is shown by dashed lines 
in  Fig.~\ref{asc}; it faces the danger of
strong coupling regime in 
either distant past or distant future, or both.

\section{An example of the bounce with conventional asymptotics}
\label{Bounce}
  
Let us now take advantage of the safety of $\gamma$-crossing and 
construct a bouncing solution in beyond Horndeski theory, where 
the driving field $\pi$
reduces to a conventional massless scalar field and gravity tends 
to GR in {\itshape both} distant past and future. 

Without loss of generality we choose the following form of the scalar
field
\[
\label{rolling_pi}
\pi (t) = t,
\]
so that $X=1$. Indeed, assuming that the scalar field monotonously 
increases, one can always obtain~\eqref{rolling_pi} by field redefinition.
Then the asymptotics of the Lagrangian functions as 
$t \rightarrow \pm\infty$ are (we set $M_{Pl}^2/(8\pi)=1$)
\begin{subequations}
\label{lagr_func_asymp}
\begin{align}
\label{F_asymp}
& F(\pi, X) = \frac{X}{3\pi^2} = \frac{1}{3t^2}, \\
\label{G4_asymp}
& G_4(\pi, X) = \frac12,\\
\label{F4_asymp}
& G_5(\pi, X) = F_4(\pi, X) = F_5(\pi, X) = 0.
\end{align}
\end{subequations}
Equations~\eqref{G4_asymp} and~\eqref{F4_asymp} ensure that gravity is
described by GR, while the choice~\eqref{F_asymp} 
indeed implies that 
$\varphi = \sqrt{\frac23}\log(\pi)$ is a conventional massless scalar 
field. Its equation of state is $p=\rho$, and hence
\[
\label{H_asymp}
H = \frac{1}{3t}, \;\; t \rightarrow \pm \infty.
\]
Note that the field equation $\ddot{\varphi} + 3H\dot{\varphi} =0$ 
is satisfied for $\varphi = \sqrt{\frac23}\log(t)$.
In this section we choose a specific form of $H$ and reconstruct 
the Lagrangian functions of the beyond Horndeski theory which yield 
the chosen solution. This approach is by now standard~\cite{Ijjas:2016tpn, Libanov:2016kfc, Kolevatov:2017voe, Cai:2017dyi}. 
Our main concern is the stability of the solution during 
the entire evolution.

Let us choose the following form of the Hubble parameter,
\[
\label{Hubble_choice}
H(t) = \frac{t}{3(\tau^2+t^2)},
\]
so that 
\[
\label{scale_factor}
a(t) = (\tau^2+t^2)^{\frac16},
\] 
and the bounce occurs at $t=0$.
The parameter $\tau$ in~\eqref{Hubble_choice} determines the duration 
of the bouncing stage; we take $\tau \gg 1$, so that the time scale
inherent in the solution greatly exceeds the Planck time.
To reconstruct the theory which admits the solution~\eqref{Hubble_choice} we use the following Ansatz for the Lagrangian functions
\begin{subequations}
\label{lagr_func_choice}
\begin{align}
\label{F}
& F(\pi, X) = f_0(\pi) + f_1(\pi)\cdot X + f_2(\pi)\cdot X^2, \\
\label{G4}
& G_4(\pi, X) = \frac12 + g_{40}(\pi) + g_{41}(\pi) \cdot X,\\
\label{F4}
& F_4(\pi, X) = f_{40}(\pi) + f_{41}(\pi) \cdot X,
\end{align}
\end{subequations} 
while $K(\pi, X) = 0$, $G_5(\pi, X) = 0$, $F_5(\pi, X) = 0$. 
Let us note that in full analogy with Ref.~\cite{Kolevatov:2017voe} there is no need 
to employ both beyond Horndeski functions $F_4(\pi, X)$ and $F_5(\pi, X)$: 
one of these functions, $F_4(\pi, X)$ in our case, is sufficient 
to get around the no-go theorem and satisfy the stability conditions. 

Our tactics is to choose $F_4(\pi, X)$, $G_4(\pi, X)$ and also $f_2(\pi)$ 
in~\eqref{lagr_func_choice} in such a way that the stability conditions are satisfied, 
and find $f_0(\pi)$ and $f_1(\pi)$ from the 
background equations of motion.
Indeed, there are two independent field equations which can be chosen, 
e.g. as $(00)$- and $(ij)$-components of the generalized Einstein 
equations (see Appendix B for their explicit forms). 
These equations can be used to find 
$f_0(\pi)$ and $f_1(\pi)$ in terms of other functions 
in~\eqref{lagr_func_choice}.
Once our $G_4(\pi, X)$ and $F_4(\pi, X)$ have 
the asymptotics~\eqref{G4_asymp},~\eqref{F4_asymp} and 
the Hubble parameter asymptotes~\eqref{H_asymp},
the function $F(\pi, X)$ automatically has the 
asymtotics~\eqref{F_asymp}.

To clarify the reasons behind further choice of 
functions in~\eqref{lagr_func_choice}, let us give an explicit form 
of the quadratic action in beyond Horndeski theory, which includes 
both tensor and scalar dynamical degrees of freedom 
(with unitary gauge imposed):  
\[
\label{action_no-go}
\begin{aligned}
S^{(2)} = \int &\mathrm{d}t\,\mathrm{d}^3x\,a^3 \Bigg[\dfrac{A_5}{2}\left(\dot{h}^T_{ik}\right)^2 - A_2 \dfrac{\left(\overrightarrow{\nabla} h_{ik}^T\right)^2}{a^2} + 
\mathcal{A} \cdot \dot{\zeta}^2 -
\mathcal{C} \cdot \dfrac{(\overrightarrow{\nabla}\zeta)^2}{a^2}
\Bigg]
\end{aligned},
\] 
where 
$\mathcal{A}$ and $\mathcal{C}$ are defined in eq.~\eqref{mathcal_ABC}.
We present the detailed reconstruction of the Lagrangian functions 
in Appendix C, and here we give the results only. 
The functions $f_0(t)$, $f_1(t)$, $f_2(t)$, $g_{40}(t)$, $g_{41}(t)$,
$f_{40}(t)$ and $f_{41}(t)$ entering~\eqref{lagr_func_choice}
are shown in Fig.~\ref{LagrangianFunctions} (the analytical expressions are gathered in Appendix C).
Their asymptotic behavior as $t \rightarrow \pm \infty$
is as follows:
\[
f_1(t) = \frac{1}{3 t^2},  \;\;f_0(t) = f_2(t) \propto \frac{1}{t^4}, \;\; g_{40}(t) = f_{40}(t) = f_{41}(t) \propto e^{-2t/\tau}, \;\; g_{41}(t) \propto t\cdot e^{-2t/\tau}.
\]
As promised, the Lagrangian functions $F(\pi,X)$, $G_4(\pi,X)$ 
and $F_4(\pi,X)$ have the asymptotics~\eqref{lagr_func_asymp}.
%
%
%
%
\begin{figure}[H]\begin{center}\hspace{-1cm}
{\includegraphics[width=0.5\linewidth]{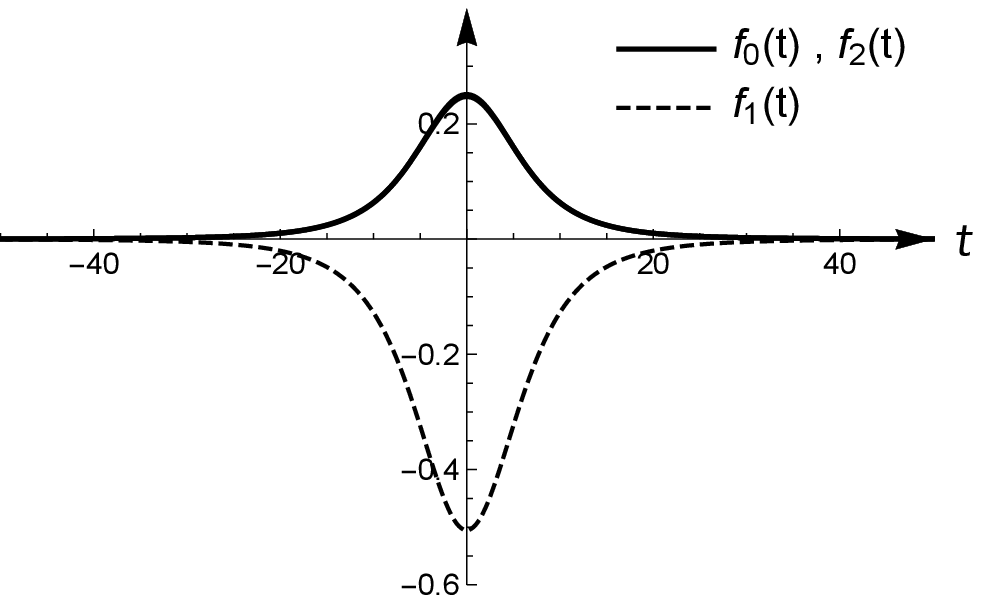}}\hspace{2.8cm}\hspace{-3cm}
{\includegraphics[width=0.5\linewidth] {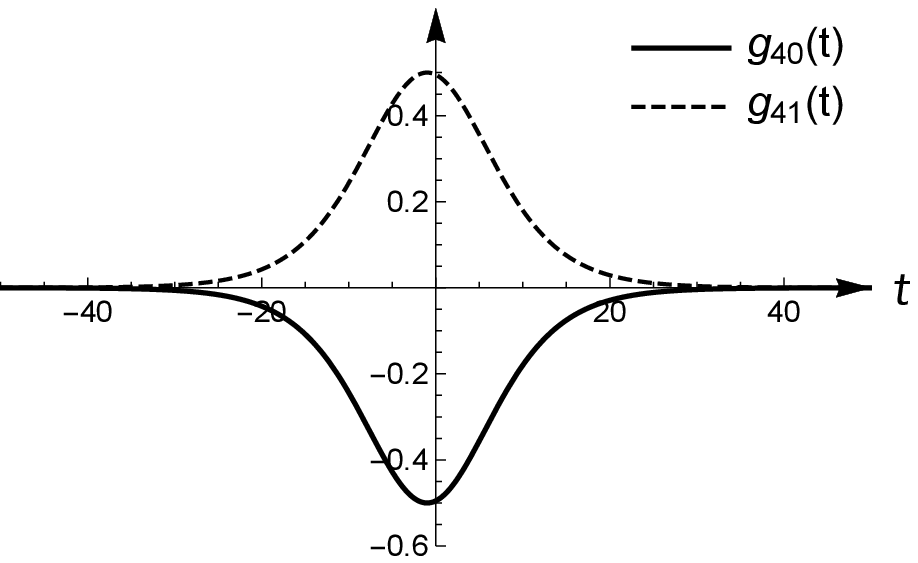}}

\vspace{-2cm}
{\includegraphics[width=0.5\linewidth]
{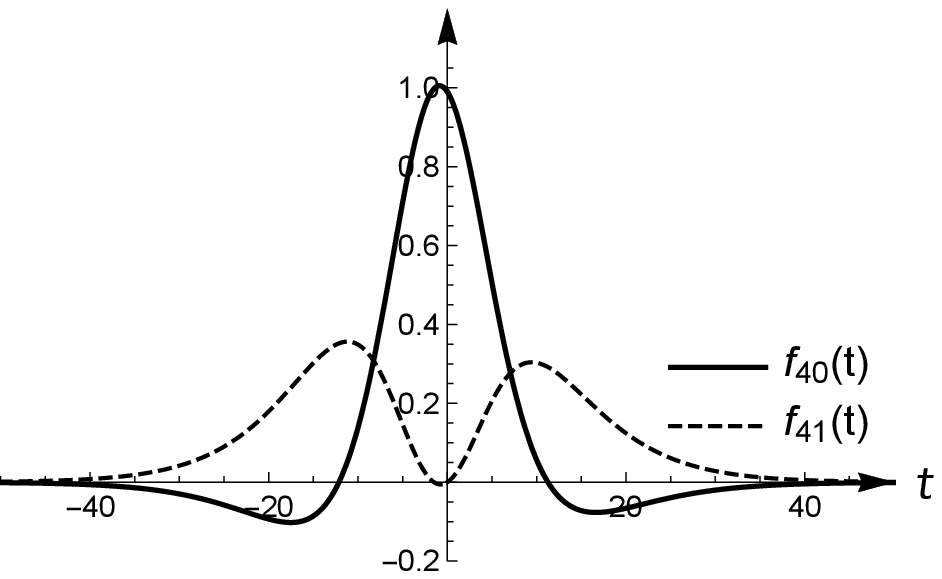}}\hspace{1cm}
\caption{\footnotesize{The Lagrangian functions $f_0(t)$, $f_1(t)$, $f_2(t)$, $g_{40}(t)$, $g_{41}(t)$,
$f_{40}(t)$ and $f_{41}(t)$, with the following choice of the parameters involved in the analytical expressions (see Appendix C): $u=1/10$, $w=1$ and $\tau = 10$. This choice guarantees that the bouncing solution is not fine-tuned and its duration safely exceeds the Planck time. Note that the functions $f_0(t)$ and $f_2(t)$ almost coincide for the chosen values of parameters.}} \label{LagrangianFunctions}
\end{center}\end{figure}
The coefficients $\mathcal{A}$ and $\mathcal{C}$ are shown in Fig.~\ref{plot_AC};
note that they are positive everywhere and infinite at 
some point ($\gamma$-crossing). 
Their ratio $c^2_{\mathcal{S}} = \mathcal{C}/\mathcal{A}$ (sound speed 
squared) is given in Fig.~\ref{sound_speed}, which shows that the propagation is 
subluminal in the scalar sector. We choose the functions $g_{40}(\pi)$,
$g_{41}(\pi)$, $f_{40}(\pi)$ and $f_{41}(\pi)$ in~\eqref{G4} and~\eqref{F4}
in such a way that 
\[
A_2 =1, \quad A_5 =2,
\]
hence, the tensor perturbations are stable and strictly luminal.

Thus, the stability requirements in both tensor and scalar sectors 
are satisfied and our bouncing solution indeed has conventional
asymptotics in both distant past and future.
\begin{figure}[h]
\center{\includegraphics[width=0.9\linewidth]{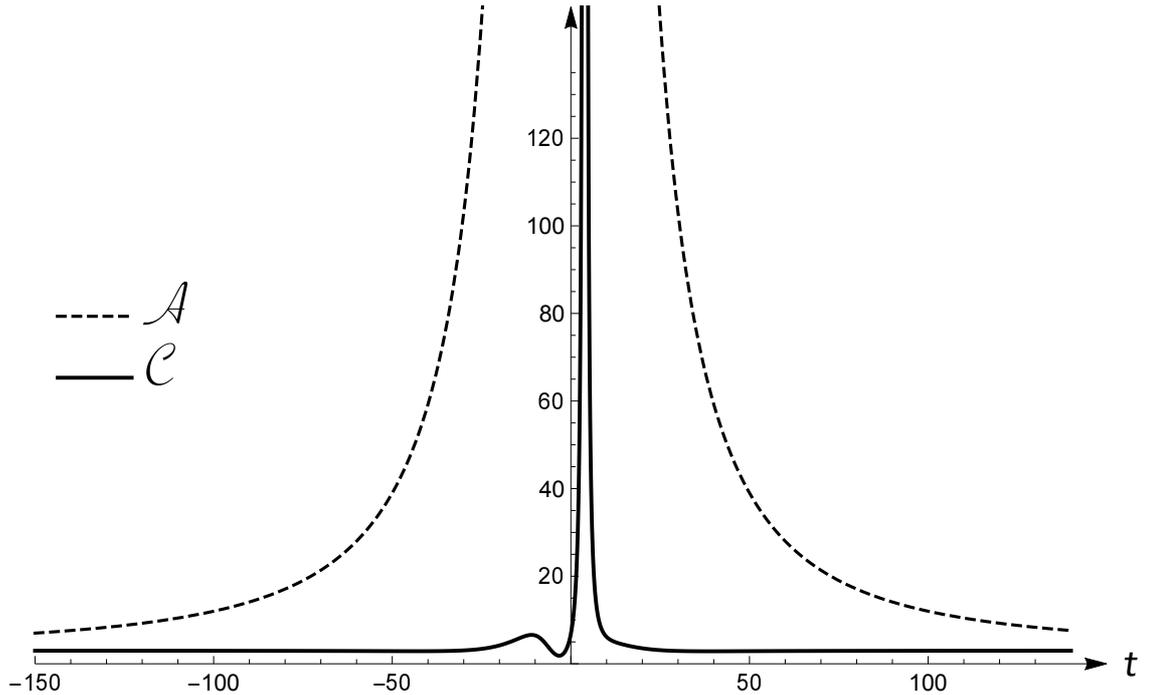} }
\caption{The coefficients $\mathcal{A}$ and $\mathcal{C}$; the parameters $u$, $w$ and $\tau$ are the same as in Fig.~\ref{LagrangianFunctions}.}
\label{plot_AC}
\end{figure}

\begin{figure}[H]
\center{\includegraphics[width=0.64\linewidth]{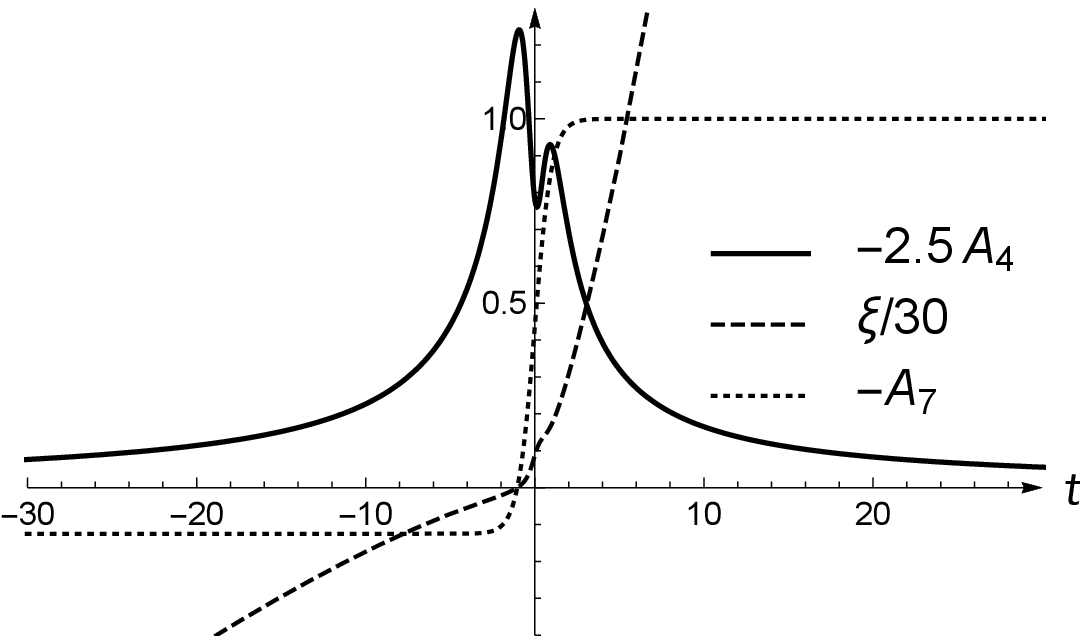}

\vspace{0.2cm}\includegraphics[width=0.64\linewidth]{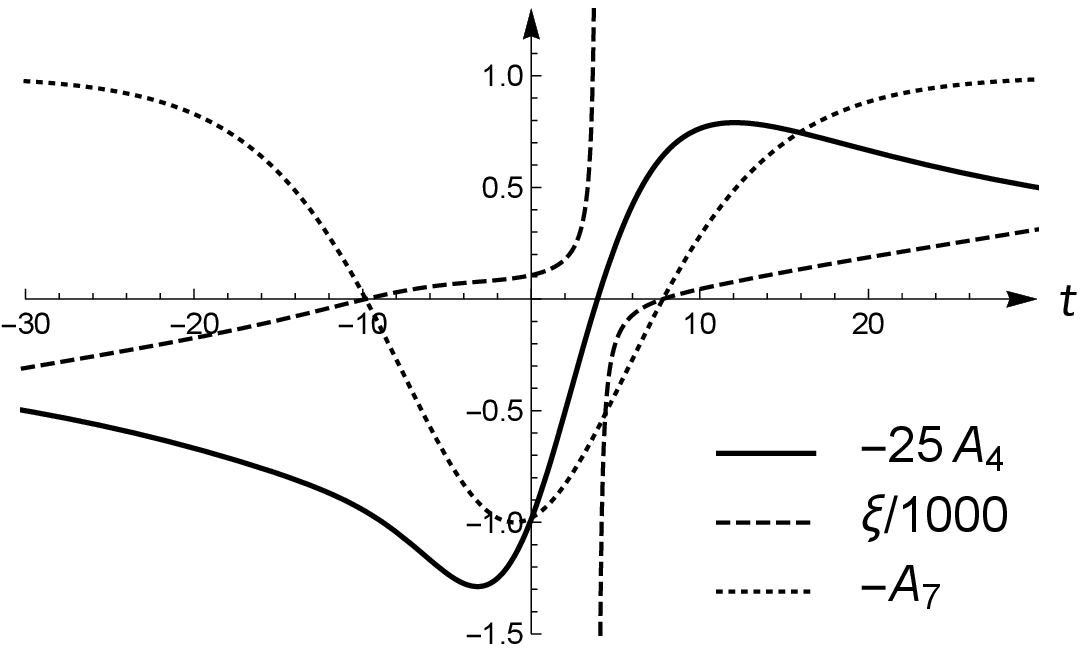} }
\caption{The coefficients $ A_4$, $A_7$ and $\xi$ for two scenarios: without $\gamma$-crossing (top panel) and with $\gamma$-crossing (bottom panel). The former case was studied in Ref.~\cite{Kolevatov:2017voe}.}
\label{nogo}
\end{figure}

Finally, let us compare bouncing models with and without 
$\gamma$-crossing. The inequality~\eqref{new_no-go} must be
satisfied in beyond Horndeski theory, so the function $\xi(t)$
defined in~\eqref{ksi} must grow monotonously. The difference,
as compared to the Horndeski theory, is that eq.~\eqref{A7toA5} does
not hold any more, so $A_7$ is no longer constrained. 
In a model without $\gamma$-crossing,
$A_4$ is always positive, so $\xi(t)$ crosses zero due to the zero of 
$A_7(t)$. This situation is shown in Fig.~\ref{nogo}, top panel.
The fact that $(-A_7)$ is negative at early times while $A_5$
is always positive (see~\eqref{action_tensor}) reiterates that 
the beyond Horndeski term is relevant at early times. 
In a model with $\gamma$-crossing, $\xi$ diverges at 
$\gamma$-crossing, so $\xi$, and hence $A_7$, crosses zero twice.
This enables one to have $-A_7 = A_5 = 1$ both at early and late 
times, which corresponds to GR asymptotics. This case is shown in 
Fig.~\ref{nogo}, lower panel.
\begin{figure}[h]
\center{\hspace{-1cm}\includegraphics[width=0.5\linewidth]{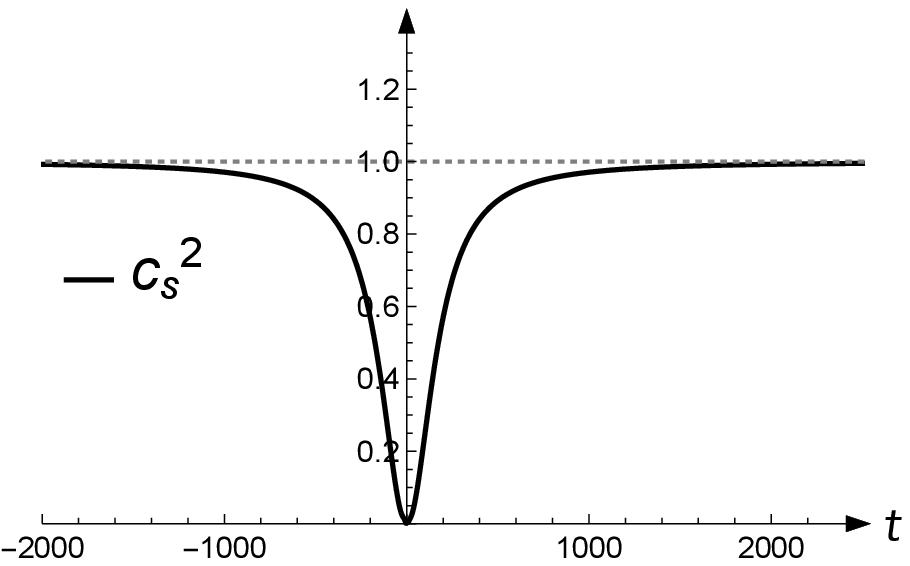}\hspace{1cm}\includegraphics[width=0.4\linewidth]{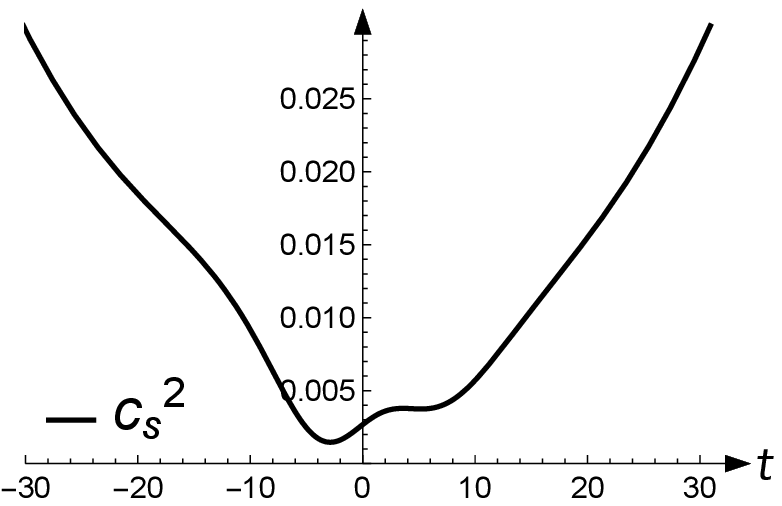} }
\caption{Sound speed squared for the scalar perturbations is non-negative for all times and asymptotically tends to 1 in both infinite past and future. Right panel shows the vicinity of the bounce. The parameters $u$, $w$, $\tau$ are the same as in Fig.~\ref{LagrangianFunctions}.}
\label{sound_speed}
\end{figure}

\section*{Acknowledgements}
The authors are grateful to Eugeny Babichev and Alexander Vikman for useful comments and fruitful discussions.
This work has been supported by Russian Science Foundation grant 14-22-00161.

\section*{Appendix A}
In this Appendix we collect the expressions for coefficients $A_i$ 
entering the quadratic action~\eqref{full_quadr_action}:
\begin{flalign}
\label{A_1}
&A_1=3\left[-2G_4+4G_{4X}\dot{\pi}^2-G_{5\pi}\dot{\pi}^2+2HG_{5X}\dot{\pi}^3 + 2 F_{4}\dot{\pi}^4\ + 6 H F_{5}\dot{\pi}^5\right],&
\end{flalign}
\vspace{-1cm}
\begin{flalign}
\label{A_2}
&A_2=2G_4-2G_{5X}\dot{\pi}^2\ddot{\pi}-G_{5\pi}\dot{\pi}^2,&
\end{flalign}
\vspace{-1cm}
\begin{flalign}\nonumber
\label{A_3}
&A_3=F_X\dot{\pi}^2+2F_{XX}\dot{\pi}^4+12HK_X\dot{\pi}^3+6HK_{XX}\dot{\pi}^5-K_{\pi}\dot{\pi}^2-K_{\pi X}\dot{\pi}^4&\\\nonumber
&-6H^2G_4+42H^2G_{4X}\dot{\pi}^2+96H^2G_{4XX}\dot{\pi}^4+24H^2G_{4XXX}\dot{\pi}^6&\\
&-6HG_{4\pi}\dot{\pi}-30HG_{4\pi X}\dot{\pi}^3-12HG_{4\pi XX}\dot{\pi}^5+30H^3G_{5X}\dot{\pi}^3&\\\nonumber
&+26H^3G_{5XX}\dot{\pi}^5+4H^3G_{5XXX}\dot{\pi}^7-18H^2G_{5\pi}\dot{\pi}^2-27H^2G_{5\pi X}\dot{\pi}^4&\\\nonumber
&-6H^2G_{5\pi XX}\dot{\pi}^6+90H^2F_4\dot{\pi}^4+78H^2F_{4X}\dot{\pi}^6+12H^2F_{4XX}\dot{\pi}^8&\\\nonumber
&+168H^3F_5\dot{\pi}^5+102H^3F_{5X}\dot{\pi}^7+12H^3F_{5XX}\dot{\pi}^9,&
\end{flalign}
\vspace{-1cm}
\begin{flalign}\nonumber
\label{A_4}
&A_4=2\big[K_X\dot{\pi}^3-2G_4H+8HG_{4X}\dot{\pi}^2+8HG_{4XX}\dot{\pi}^4-G_{4\pi}\dot{\pi}-2G_{4\pi X}\dot{\pi}^3&\\
&+5H^2G_{5X}\dot{\pi}^3+2H^2G_{5XX}\dot{\pi}^5-3HG_{5\pi}\dot{\pi}^2-2HG_{5\pi X}\dot{\pi}^4&\\\nonumber
&+10HF_4\dot{\pi}^4+4HF_{4X}\dot{\pi}^6+21H^2F_5\dot{\pi}^5+6H^2F_{5X}\dot{\pi}^7\big],&
\end{flalign}
\vspace{-1cm}
\begin{flalign}
\label{A_5}
&A_5=-\dfrac{2}{3}A_1,&
\end{flalign}
\vspace{-1cm}
\begin{flalign}
\label{A_6}
&A_6=-3A_4,&
\end{flalign}
\vspace{-1cm}
\begin{flalign}
\label{A_7}
&A_7 = -A_5 -B_{16}\dot{\pi},&
\end{flalign}
\vspace{-1cm}
\begin{flalign}\nonumber
\label{A_8}
&A_8=2\big[K_X\dot{\pi}^2-G_{4\pi}-2G_{4\pi X}\dot{\pi}^2+4 HG_{4X}\dot{\pi}+8HG_{4XX}\dot{\pi}^3-2HG_{5\pi}\dot{\pi}&\\
&-2HG_{5\pi X}\dot{\pi}^3+3 H^2G_{5X}\dot{\pi}^2+2 H^2G_{5XX}\dot{\pi}^4 + 10 H F_{4}\dot{\pi}^3
+ 4 H F_{4X}\dot{\pi}^5 &\\\nonumber
&+ 21 H^2 F_{5}\dot{\pi}^4 + 6 H^2 F_{5X}\dot{\pi}^6\big],&
\end{flalign}
\vspace{-1cm}
\begin{flalign}
\label{A_9}
&A_9=- \big(A_8 - B_{16} H\big),&
\end{flalign}
\vspace{-1cm}
\begin{flalign}
\label{A_10}
&A_{10}=-3\big(A_8-B_{16} H\big) ,&
\end{flalign}
\vspace{-1cm}
\begin{flalign}\nonumber
\label{A_11}
&A_{11}=2\big[-F_X\dot{\pi}-2F_{XX}\dot{\pi}^3+K_\pi\dot{\pi}-6HK_{XX}\dot{\pi}^4-9HK_X\dot{\pi}^2+K_{\pi X}\dot{\pi}^3&\\\nonumber
&+3HG_{4\pi}+24HG_{4\pi X}\dot{\pi}^2+12H G_{4\pi XX}\dot{\pi}^4-18H^2G_{4X}\dot{\pi}-72H^2G_{4XX}\dot{\pi}^3&\\
&-24H^2G_{4XXX}\dot{\pi}^5+9H^2G_{5\pi}\dot{\pi}+21H^2G_{5\pi X}\dot{\pi}^3+6H^2G_{5\pi XX}\dot{\pi}^5&\\\nonumber
&-15H^3G_{5X}\dot{\pi}^2-
20H^3G_{5XX}\dot{\pi}^4-4H^3G_{5XXX}\dot{\pi}^6 - 60 H^2 F_{4}\dot{\pi}^3 - 66 H^2 F_{4X}\dot{\pi}^5&\\\nonumber
& - 12 H^2 F_{4XX}\dot{\pi}^7 - 105 H^3 F_{5}\dot{\pi}^4 - 84 H^3 F_{5X}\dot{\pi}^6 - 12 H^3 F_{5XX} \dot{\pi}^8
\big],&
\end{flalign}
\vspace{-1cm}
\begin{flalign}\nonumber
\label{A_12}
&A_{12}=2\big[F_X\dot{\pi}-K_\pi\dot{\pi}+3HK_X\dot{\pi}^2-HG_{4\pi}+G_{4\pi\pi}\dot{\pi}-10HG_{4\pi X}\dot{\pi}^2+6H^2G_{4X}\dot{\pi}&\\
&+12H^2G_{4XX}\dot{\pi}^3-3H^2G_{5\pi}\dot{\pi}+HG_{5\pi\pi}\dot{\pi}^2-4H^2G_{5\pi X}\dot{\pi}^3+3H^3G_{5X}\dot{\pi}^2&\\\nonumber
&+2H^3G_{5XX}\dot{\pi}^4 + 12 H^2 F_{4}\dot{\pi}^3 + 6 H^2 F_{4X}\dot{\pi}^5 - 2 H F_{4\pi}\dot{\pi}^4 + 15 H^3 F_{5} \dot{\pi}^4 + 6 H^3 F_{5X} \dot{\pi}^6 &\\\nonumber
&- 3 H^2 F_{5\pi} \dot{\pi}^5
\big],&
\end{flalign}
\vspace{-1cm}
\begin{flalign}\nonumber
\label{A_13}
&A_{13}= 2\big[4HG_{4X}\dot{\pi} + 4G_{4X}\ddot{\pi} + 8 G_{4XX}\dot{\pi}^2\ddot{\pi} - 2G_{4\pi} +4 G_{4\pi X}\dot{\pi}^2 + 2 H^2G_{5X}\dot{\pi}^2&\\\nonumber
&+2 \dot{H}G_{5X}\dot{\pi}^2+4 H G_{5X}\dot{\pi}\ddot{\pi}+4HG_{5XX}\dot{\pi}^3\ddot{\pi} -2HG_{5\pi}\dot{\pi}-2G_{5\pi}\ddot{\pi}+2HG_{5\pi X}\dot{\pi}^3&\\
&-2G_{5\pi X}\dot{\pi}^2\ddot{\pi}-G_{5\pi\pi}\dot{\pi}^2 + 2 H F_{4}\dot{\pi}^3 + 6 F_{4} \ddot{\pi}\dot{\pi}^2 +4 F_{4X}\ddot{\pi}\dot{\pi}^4 + 2 F_{4\pi}\dot{\pi}^4 + 24 H F_{5} \ddot{\pi}\dot{\pi}^3&\\\nonumber
&+6 H^2 F_{5} \dot{\pi}^4 + 6 \dot{H} F_{5}\dot{\pi}^4 + 12 H F_{5X}\ddot{\pi}\dot{\pi}^5 + 6 H F_{5\pi} \dot{\pi}^5
\big],
\end{flalign}
\vspace{-1cm}
\begin{flalign}\nonumber
\label{A_14}
&A_{14}=F_{X} + 2 F_{XX}\dot{\pi}^2- K_{\pi} + 6 H K_{X}\dot{\pi} -
 K_{\pi X}\dot{\pi}^2 + 6 H K_{XX}\dot{\pi}^3 + 6 H^2 G_{4 X} &\\\nonumber
 &-18 H G_{4\pi X}\dot{\pi} + 48 H^2 G_{4 XX}\dot{\pi}^2 -
 12 H G_{4\pi XX}\dot{\pi}^3 + 24 H^2 G_{4 XXX}\dot{\pi}^4 +
 6 H^3 G_{5 X}\dot{\pi} &\\
 &- 3 H^2 G_{5\pi}-
15 H^2 G_{5\pi X}\dot{\pi}^2 + 14 H^3 G_{5 XX}\dot{\pi}^3 +
 4 H^3 G_{5 XXX}\dot{\pi}^5 - 6 H^2 G_{5\pi XX}\dot{\pi}^4 &\\\nonumber
 &+
 36 H^2 F_{4}\dot{\pi}^2 + 54 H^2 F_{4 X}\dot{\pi}^4 +
 12 H^2 F_{4 XX}\dot{\pi}^6 + 60 H^3 F_{5}\dot{\pi}^3 +
 66 H^3 F_{5 X}\dot{\pi}^5 &\\\nonumber
 &+ 12 H^3 F_{5 XX}\dot{\pi}^7 ,&
\end{flalign}
\vspace{-1cm}
\begin{flalign}\nonumber
\label{A_15}
&A_{15}=-F_{X} - 4 H K_{X}\dot{\pi} - 2 K_{X}\ddot{\pi} + K_{\pi} -
 K_{\pi X}\dot{\pi}^2 - 2  K_{XX}\dot{\pi}^2 \ddot{\pi} -
 6 H^2 G_{4 X}& \\\nonumber
 &- 4 \dot{H} G_{4 X} -
 20 H^2 G_{4 XX}\dot{\pi}^2 - 8 \dot{H} G_{4 XX}\dot{\pi}^2 -
 24 H  G_{4 XX}\dot{\pi}\ddot{\pi} + 12 H  G_{4\pi X}\dot{\pi} \\\nonumber
 &+
 6 G_{4\pi X}\ddot{\pi} - 16 H  G_{4 XXX}\dot{\pi}^3\ddot{\pi} -
 8 H G_{4\pi XX}\dot{\pi}^3 + 4  G_{4\pi XX}\dot{\pi}^2\ddot{\pi} +
 2  G_{4\pi \pi X}\dot{\pi}^2& \\\nonumber
 &- 4 H^3 G_{5 X}\dot{\pi} -
 4 H\dot{H} G_{5 X}\dot{\pi} - 2 H^2  G_{5 X}\ddot{\pi} +
 3 H^2 G_{5\pi} + 2 \dot{H} G_{5\pi} +
 5 H^2  G_{5\pi X}\dot{\pi}^2 &\\
 &+
 2 \dot{H} G_{5\pi X}\dot{\pi}^2 +
 8 H  G_{5\pi X}\dot{\pi}\ddot{\pi} -
 4 H^3  G_{5 XX}\dot{\pi}^3 - 4 H \dot{H} G_{5 XX}\dot{\pi}^3 -
 10 H^2  G_{5 XX}\dot{\pi}^2\ddot{\pi}& \\\nonumber
 &-
 4 H^2  G_{5 XXX}\dot{\pi}^4\ddot{\pi} -
 2 H^2  G_{5\pi XX}\dot{\pi}^4 +
 4 H  G_{5\pi XX}\dot{\pi}^3\ddot{\pi} +
 2 H  G_{5\pi\pi X}\dot{\pi}^3 - 20  F_{4} H^2 \dot{\pi}^2& \\\nonumber
 &-
 10 \dot{H} F_{4}  \dot{\pi}^2 -
 24 H F_{4}\dot{\pi} \ddot{\pi} - 10 H^2  F_{4 X}\dot{\pi}^4 -
 4 \dot{H} F_{4 X}\dot{\pi}^4 -
 36 H  F_{4 X}\dot{\pi}^3\ddot{\pi} - 6 H  F_{4\pi}\dot{\pi}^3& \\\nonumber
 &-
 8 H  F_{4 XX}\dot{\pi}^5\ddot{\pi} -
 4 H  F_{4\pi X}\dot{\pi}^5 - 30 H^3 F_{5} \dot{\pi}^3 -
 36 H\dot{H} F_{5} \dot{\pi}^3 -
 60 H^2 F_{5} \dot{\pi}^2\ddot{\pi} -
 12 H^3 F_{5 X}\dot{\pi}^5& \\\nonumber
 &- 12 H \dot{H} F_{5 X}\dot{\pi}^5 -
 66 H^2  F_{5 X}\dot{\pi}^4\ddot{\pi} -
 12 H^2 F_{5\pi}\dot{\pi}^4 -
 12 H^2  F_{5 XX}\dot{\pi}^6\ddot{\pi} -
 6 H^2  F_{5\pi X}\dot{\pi}^6,&
\end{flalign}
\vspace{-1cm}
\begin{flalign}
\label{B_16}
&B_{16}=4 F_{4} \dot{\pi}^3 + 12 H F_{5}\dot{\pi}^4,&
\end{flalign}
\vspace{-1cm}
\begin{flalign}\nonumber
&A_{17}=F_{\pi}-2 F_{\pi X} \dot{\pi}^2+K_{\pi\pi{}} \dot{\pi}^2-6 H K_{\pi X} \dot{\pi}^3+6 G_{4\pi}  H^2+6 G_{4\pi\pi{}} H \dot{\pi}-24 G_{4\pi X} H^2 \dot{\pi}^2 \\
&+12 G_{4\pi\pi{}X} H \dot{\pi}^3-24 G_{4\pi XX} H^2 \dot{\pi}^4+9 G_{5\pi\pi{}} H^2 \dot{\pi}^2-10 G_{5\pi X} H^3 \dot{\pi}^3+6 G_{5\pi\pi{}X} H^2 \dot{\pi}^4\\\nonumber
&-4 G_{5\pi XX} H^3 \dot{\pi}^5-30 F_{4\pi} H^2 \dot{\pi}^4-12 F_{4\pi X} H^2 \dot{\pi}^6-42 F_{5\pi} H^3 \dot{\pi}^5-12 F_{5\pi X} H^3 \dot{\pi}^7,&
\end{flalign}
\begin{flalign}\nonumber
&A_{18}=-6 F_{X} \dot{\pi}+6 K_{\pi} \dot{\pi}-36 H K_{X} \dot{\pi}^2-12 K_{X} \dot{\pi} \ddot{\pi}-6 K_{\pi X} \dot{\pi}^3-12 K_{XX} \dot{\pi}^3 \ddot{\pi}+24 G_{4\pi} H \\\nonumber
&-24 G_{4X} H \ddot{\pi}-108 G_{4X} H^2 \dot{\pi}-24 G_{4X} \dot{H} \dot{\pi}+72 G_{4\pi X} H \dot{\pi}^2-216 G_{4XX} H^2 \dot{\pi}^3\\\nonumber
&-48 G_{4XX} \dot{H} \dot{\pi}^3+36 G_{4\pi X} \dot{\pi} \ddot{\pi}-192 G_{4XX} H \dot{\pi}^2 \ddot{\pi}+24 G_{4\pi XX} \dot{\pi}^3 \ddot{\pi}-96 G_{4XXX} H \dot{\pi}^4 \ddot{\pi}\\\nonumber
&-48 G_{4\pi XX} H \dot{\pi}^4+12 G_{4\pi\pi{}X} \dot{\pi}^3+54 G_{5\pi} H^2 \dot{\pi}+12 G_{5\pi} \dot{H} \dot{\pi}+12 G_{5\pi} H \ddot{\pi}-36 G_{5X} H^2 \dot{\pi} \ddot{\pi}\\\nonumber
&-72 G_{5X} H^3 \dot{\pi}^2-36 G_{5X} H \dot{H} \dot{\pi}^2+6 G_{5\pi\pi{}} H \dot{\pi}^2+42 G_{5\pi X} H^2 \dot{\pi}^3+12 G_{5\pi X} \dot{H} \dot{\pi}^3\\
&+60 G_{5\pi X} H \dot{\pi}^2 \ddot{\pi}-84 G_{5XX} H^2 \dot{\pi}^3 \ddot{\pi}-48 G_{5XX} H^3 \dot{\pi}^4-24 G_{5XX} H \dot{H} \dot{\pi}^4\\\nonumber
&+12 G_{5\pi\pi{}X} H \dot{\pi}^4-12 G_{5\pi XX} H^2 \dot{\pi}^5+24 G_{5\pi XX} H \dot{\pi}^4 \ddot{\pi}-24 G_{5XXX} H^2 \dot{\pi}^5 \ddot{\pi}-216 F_{4} H^2 \dot{\pi}^3\\\nonumber
&-48 F_{4} \dot{H} \dot{\pi}^3-144 F_{4} H \dot{\pi}^2 \ddot{\pi}-36 F_{4\pi} H \dot{\pi}^4-108 F_{4X} H^2 \dot{\pi}^5-24 F_{4X} \dot{H} \dot{\pi}^5-216 F_{4X} H \dot{\pi}^4 \ddot{\pi}\\\nonumber
&-24 F_{4\pi X} H \dot{\pi}^6-48 F_{4XX} H \dot{\pi}^6 \ddot{\pi}-360 F_{5} H^3 \dot{\pi}^4-180 F_{5} H \dot{H} \dot{\pi}^4-360 F_{5} H^2 \dot{\pi}^3 \ddot{\pi}-72 F_{5\pi} H^2 \dot{\pi}^5\\\nonumber
&-144 F_{5X} H^3 \dot{\pi}^6-72 F_{5X} H \dot{H} \dot{\pi}^6-396 F_{5X} H^2 \dot{\pi}^5 \ddot{\pi}-36 F_{5\pi X} H^2 \dot{\pi}^7-72 F_{5\pi\pi{}} H^2 \dot{\pi}^7 \ddot{\pi},&
\end{flalign}
\begin{flalign}\nonumber
&A_{19}=3 F_{\pi}-18 F_{X} H \dot{\pi}-6 F_{X} \ddot{\pi}-6 F_{\pi X} \dot{\pi}^2-12 F_{XX} \dot{\pi}^2 \ddot{\pi}+18 H K_{\pi} \dot{\pi}+6 K_{\pi} \ddot{\pi}-54 H^2 K_{X} \dot{\pi}^2\\\nonumber
&-36 H K_{X} \dot{\pi} \ddot{\pi}-18 \dot{H} K_{X} \dot{\pi}^2-36 H K_{XX} \dot{\pi}^3 \ddot{\pi}+3 K_{\pi\pi{}} \dot{\pi}^2-18 H K_{\pi X} \dot{\pi}^3+6 K_{\pi X} \dot{\pi}^2 \ddot{\pi}\\
&+36 G_{4\pi} H^2+18 G_{4\pi} \dot{H}-108 G_{4X} H^3 \dot{\pi}-72 G_{4X} H \dot{H} \dot{\pi}-36 G_{4X} H^2 \ddot{\pi}+108 G_{4\pi X} H^2 \dot{\pi}^2\\\nonumber
&+36 G_{4\pi X} \dot{H} \dot{\pi}^2+108 G_{4\pi X} H \dot{\pi} \ddot{\pi}-288 G_{4XX} H^2 \dot{\pi}^2 \ddot{\pi}-216 G_{4XX} H^3 \dot{\pi}^3-144 G_{4XX} H \dot{H} \dot{\pi}^3\\\nonumber
&-72 G_{4\pi XX} H^2 \dot{\pi}^4+36 G_{4\pi\pi{}X} H \dot{\pi}^3+72 G_{4\pi XX} H \dot{\pi}^3 \ddot{\pi}-144 G_{4XXX} H^2 \dot{\pi}^4 \ddot{\pi}+54 G_{5\pi} H^3 \dot{\pi}\\\nonumber
&+36 G_{5\pi} H \dot{H} \dot{\pi}+18 G_{5\pi} H^2 \ddot{\pi}-36 G_{5X} H^3 \dot{\pi} \ddot{\pi}-54 G_{5X} H^4 \dot{\pi}^2-54 G_{5X} H^2 \dot{H} \dot{\pi}^2+9 G_{5\pi\pi{}} H^2 \dot{\pi}^2\\\nonumber
&+42 G_{5\pi X} H^3 \dot{\pi}^3+36 G_{5\pi X} H \dot{H} +90 G_{5\pi X} H^2 \dot{\pi}^2 \ddot{\pi}-84 G_{5XX} H^3 \dot{\pi}^3 \ddot{\pi}-36 G_{5XX} H^4 \dot{\pi}^4\\\nonumber
&-36 G_{5XX} H^2 \dot{H} \dot{\pi}^4 \dot{\pi}^3+18 G_{5\pi\pi{}X} H^2 \dot{\pi}^4-12 G_{5\pi XX} H^3 \dot{\pi}^5+36 G_{5\pi XX} H^2 \dot{\pi}^4 \ddot{\pi}-24 G_{5XXX} H^3 \dot{\pi}^5 \ddot{\pi}\\\nonumber
&-216 F_{4} H^3 \dot{\pi}^3-144 F_{4} H \dot{H} \dot{\pi}^3-216 F_{4} H^2 \dot{\pi}^2 \ddot{\pi}-54 F_{4\pi} H^2 \dot{\pi}^4-108 F_{4X} H^3 \dot{\pi}^5-72 F_{4X} H \dot{H} \dot{\pi}^5\\\nonumber
&-324 F_{4X} H^2 \dot{\pi}^4 \ddot{\pi}-36 F_{4\pi X} H^2 \dot{\pi}^6-72 F_{4XX} H^2 \dot{\pi}^6 \ddot{\pi}-270 F_{5} H^4 \dot{\pi}^4-270 F_{5} H^2 \dot{H} \dot{\pi}^4\\\nonumber
&-360 F_{5} H^3 \dot{\pi}^3 \ddot{\pi}-72 F_{5\pi} H^3 \dot{\pi}^5-108 F_{5X} H^4 \dot{\pi}^6-108 F_{5X} H^2 \dot{H} \dot{\pi}^6-396 F_{5X} H^3 \dot{\pi}^5 \ddot{\pi}\\\nonumber
&-36 F_{5\pi X} H^3 \dot{\pi}^7-72 F_{5\pi\pi{}} H^3 \dot{\pi}^7 \ddot{\pi},&
\end{flalign}
\begin{flalign}\nonumber
&A_{20}=\frac12 F_{\pi\pi{}}-3 F_{\pi X} H \dot{\pi}-F_{\pi X} \ddot{\pi}-F_{\pi\pi{}X} \dot{\pi}^2-2 F_{\pi XX} \dot{\pi}^2 \ddot{\pi}-9 H^2 K_{\pi X} \dot{\pi}^2-3 \dot{H} K_{\pi X} \dot{\pi}^2\\\nonumber
&-6 H K_{\pi X} \dot{\pi} \ddot{\pi}+3 H K_{\pi\pi{}} \dot{\pi}+K_{\pi\pi{}} \ddot{\pi}+\frac12 K_{\pi\pi\pi{}} \dot{\pi}^2-3 H K_{\pi\pi{}X} \dot{\pi}^3+K_{\pi\pi{}X} \dot{\pi}^2 \ddot{\pi}-6 H K_{\pi XX} \dot{\pi}^3 \ddot{\pi}\\\nonumber
&+6 G_{4\pi\pi{}} H^2+3 G_{4\pi\pi{}} \dot{H}-18 G_{4\pi X} H^3 \dot{\pi}-12 G_{4\pi X} H \dot{H} \dot{\pi}-6 G_{4\pi X} H^2 \ddot{\pi}+18 G_{4\pi\pi{}X} H^2 \dot{\pi}^2\\\nonumber
&+6 G_{4\pi\pi{}X} \dot{H} \dot{\pi}^2-36 G_{4\pi XX} H^3 \dot{\pi}^3-48 G_{4\pi XX} H^2 \dot{\pi}^2 \ddot{\pi}-24 G_{4\pi XX} H \dot{H} \dot{\pi}^3-12 G_{4\pi\pi{}XX} H^2 \dot{\pi}^4\\\nonumber
&+18 G_{4\pi\pi{}X} H \dot{\pi} \ddot{\pi}+12 G_{4\pi\pi{}XX} H \dot{\pi}^3 \ddot{\pi}-24 G_{4\pi XXX} H^2 \dot{\pi}^4 \ddot{\pi}+6 G_{4\pi\pi{}\pi X} H \dot{\pi}^3+9 G_{5\pi\pi{}} H^3 \dot{\pi}\\\nonumber
&+6 G_{5\pi\pi{} } H \dot{H} \dot{\pi}+3 G_{5\pi\pi{}} H^2 \ddot{\pi}-6 G_{5\pi X} H^3 \dot{\pi} \ddot{\pi}-9 G_{5\pi X} H^4 \dot{\pi}^2-9 G_{5\pi X} H^2 \dot{H} \dot{\pi}^2+\frac32 G_{5\pi\pi{}\pi} H^2 \dot{\pi}^2\\\nonumber
&+7 G_{5\pi\pi{}X} H^3 \dot{\pi}^3+6 G_{5\pi\pi{}X} H \dot{H} \dot{\pi}^3-6 G_{5\pi XX} H^4 \dot{\pi}^4-6 G_{5\pi XX} H^2 \dot{H} \dot{\pi}^4+15 G_{5\pi\pi{}X} H^2 \dot{\pi}^2 \ddot{\pi}\\\nonumber
&-14 G_{5\pi XX} H^3 \dot{\pi}^3 \ddot{\pi}+6 G_{5\pi\pi{}XX} H^2 \dot{\pi}^4 \ddot{\pi}-2 G_{5\pi\pi{}XX} H^3 \dot{\pi}^5+3 G_{5\pi\pi{}\pi X} H^2 \dot{\pi}^4-4 G_{5\pi XXX} H^3 \dot{\pi}^5 \ddot{\pi}\\\nonumber
&-36 F_{4\pi} H^3 \dot{\pi}^3-24 F_{4\pi} H \dot{H} \dot{\pi}^3-36 F_{4\pi} H^2 \dot{\pi}^2 \ddot{\pi}-9 F_{4\pi\pi{}} H^2 \dot{\pi}^4-18 F_{4\pi X} H^3 \dot{\pi}^5-12 F_{4\pi X} H \dot{H} \dot{\pi}^5\\\nonumber
&-54 F_{4\pi X} H^2 \dot{\pi}^4 \ddot{\pi}-6 F_{4\pi\pi{}X} H^2 \dot{\pi}^6-12 F_{4\pi XX} H^2 \dot{\pi}^6 \ddot{\pi}-45 F_{5\pi} H^4 \dot{\pi}^4-45 F_{5\pi} H^2 \dot{H} \dot{\pi}^4\\
&-60 F_{5\pi} H^3 \dot{\pi}^3 \ddot{\pi}-12 F_{5\pi\pi{}} H^3 \dot{\pi}^5-66 F_{5\pi X} H^3 \dot{\pi}^5 \ddot{\pi}-18 F_{5\pi X} H^4 \dot{\pi}^6-18 F_{5\pi X} H^2 \dot{H} \dot{\pi}^6\\\nonumber
&-6 F_{5\pi\pi{}X} H^3 \dot{\pi}^7-12 F_{5\pi XX}  H^3 \dot{\pi}^7 \ddot{\pi{}}.&
\end{flalign}
Note that $B_{16}=0$ in the general Horndeski theory.

\section*{Appendix B}
This Appendix gives the $(00)$- and $(ij)$- components of the generalized
Einstein equations for spatially flat FLRW background
in beyond Horndeski theory~\eqref{lagrangian}:
\begin{subequations}
	\label{einst_eq}
	\begin{align*}
	\delta g^{00}:\quad
	&F-2F_XX-6HK_XX\dot{\pi}+K_{\pi}X+6H^2G_4+6HG_{4\pi}\dot{\pi}
	\\\nonumber&-24H^2X(G_{4X}+G_{4XX}X)+12HG_{4\pi X}X\dot{\pi}
	\\\nonumber&-2H^3X\dot{\pi}(5G_{5X}+2G_{5XX}X)+3H^2X(3G_{5\pi}+2G_{5\pi X}X)
	\\\nonumber&-6H^2X^2(5F_4+2F_{4X}X)-6H^3X^2\dot{\pi}(7F_5+2F_{5X}X)=0,
	\\\nonumber
	\\
	\delta g^{ij}:\quad
	&F-X(2K_X\ddot{\pi}+K_\pi)+2(3H^2+2\dot{H})G_4-12H^2G_{4X}X
	\\\nonumber&-8\dot{H}G_{4X}X-8HG_{4X}\ddot{\pi}\dot{\pi}-16HG_{4XX}X\ddot{\pi}\dot{\pi}+2(\ddot{\pi}+2H\dot{\pi})G_{4\pi}
	\\\nonumber&+4XG_{4\pi X}(\ddot{\pi}-2H\dot{\pi})+2XG_{4\pi\pi}-2XG_{5X}(2H^3\dot{\pi}+2H\dot{H}\dot{\pi}+3H^2\ddot{\pi})
	\\\nonumber&-4H^2G_{5XX}X^2\ddot{\pi}+G_{5\pi}(3H^2X+2\dot{H}X+4H\ddot{\pi}\dot{\pi})
	\\\nonumber&+2HG_{5\pi X}X(2\ddot{\pi}\dot{\pi}-HX)+2HG_{5\pi\pi}X\dot{\pi}
	\\\nonumber&-2F_4X(3H^2X+2\dot{H}X+8H\ddot{\pi}\dot{\pi})-8HF_{4X}X^2\ddot{\pi}\dot{\pi}-4HF_{4\pi}X^2\dot{\pi}
	\\\nonumber&-6HF_5X^2(2H^2\dot{\pi}+2\dot{H}\dot{\pi}+5H\ddot{\pi})-12H^2F_{5X}X^3\ddot{\pi}-6H^2F_{5\pi}X^3=0,
	\end{align*}
\end{subequations}
where $X = \dot{\pi}^2$.

\section*{Appendix C}
In this Appendix we describe in detail the reconstruction procedure for 
the Lagrangian functions shown in Fig.~\ref{LagrangianFunctions}.

The procedure is based, in particular, on the form of the 
quadratic action~\eqref{action_no-go} for tensor and scalar 
perturbations in beyond Horndeski theory. As before, the coefficients 
$\mathcal{A}$ and $\mathcal{C}$ are given by~\eqref{mathcal_ABC}. 
Making use of expansions~\eqref{lagr_func_choice},
we cast $A_2$, $A_3$, $A_4$, $A_5$ and $A_7$ 
(eqs.~\eqref{A_2},~\eqref{A_3},
~\eqref{A_4},~\eqref{A_5} and ~\eqref{A_7} in Appendix A) 
in the following form:
\begin{subequations}
\begin{align}
\label{Ft}
 A_2 &= 2 \left(\frac12 + g_{40}(t) + g_{41}(t)\right),\\
\label{Sigma}
 A_3 &= f_1(t)+6 f_2(t)-3 H^2\cdot \left[1+30 f_{40}(t)+56 f_{41}(t)+2 g_{40}(t)-12 g_{41}(t)\right]\\
 \nonumber
&-6 H\cdot \left[\dot{g}_{40}(t)+6 \dot{g}_{41}(t)\right],\\
\label{Theta}
A_4 &= 
-2 \cdot H(t)\cdot\left[1 + 2 g_{40}(t) - 6 g_{41}(t) + 10 f_{40}(t) + 14 f_{41}(t)\right] - 2\dot{g}_{40}(t) - 6 \dot{g}_{41}(t),\\
\label{Gtt}
 A_5 &= 4 \left(\frac12 + g_{40}(t)+g_{41}(t)\right) - 8 g_{41}(t) + 4 \left(f_{40}(t)+f_{41}(t)\right),\\
\label{Gt}
A_7 &=  -4 \left(\frac12 + g_{40}(t)+g_{41}(t)\right) + 8 g_{41}(t).
\end{align}
\end{subequations}
We heavily rely on these expressions when choosing the
functions $g_{40}(t)$, $g_{41}(t)$, $f_{40}(t)$, $f_{41}(t)$ and $f_{2}(t)$.  

First, we require that ghost and gradient instabilities are absent 
in the tensor sector, i.e. $A_5 > 0$ and $A_2 > 0$.
According to the requirement of asymptotically vanishing 
$F_4(\pi, X)$ in~\eqref{F4_asymp}, one possible choice
for $f_{40}(t)$ in~\eqref{Gtt} is the following:
\[
\label{f_40}
f_{40}(t) = -f_{41}(t) + w\cdot \mbox{sech}^2\left(\frac{t}{\tau}+u\right),
\]
where $u$ and $w$ are constants, which are introduced to avoid fine-tuning. We give a detailed discussion of the fine-tuning 
issue below.

In order to avoid superluminal propagation of the tensor modes,
let us choose
\[
\label{A2_A5}
A_2 = 1, \quad A_5 = 2,
\]
and express $g_{41}(t)$ and $g_{40}(t)$ using eqs.~\eqref{Gtt} 
and~\eqref{Ft}, respectively,
\[
\label{g40_g41}
g_{40}(t) = - g_{41}(t) = - \frac{w}{2} \;\mbox{sech}^2\left(\frac{t}{\tau}+u\right).
\]
Thus, we have completely defined $G_4(\pi, X)$ in~\eqref{G4}. 

Let us now use the stability conditions for the scalar sector of action~\eqref{action_no-go}. To have no gradient instabilities one requires $\mathcal{C} > 0$. So according to eqs.~\eqref{mathcal_C}
\[
\label{new_ksi}
\frac{1}{a}\frac{d\xi}{dt} = 
\frac{1}{a}\frac{d}{dt}\left[ \frac{a A_5 \cdot A_7}{2 A_4}\right] > A_2
\] 
where the expression for $\xi$ given by~\eqref{ksi} is used. 
Since $A_2>0$, $\xi$ must be a monotonously growing function 
of time, so it crosses zero at some point(s). We have already made a 
choice~\eqref{A2_A5}, so we are left with $A_7$ and $A_4$ in~\eqref{new_ksi}. 
Note that, unlike in the case of the general Horndeski theory, $A_7$ 
is not constrained by any stability conditions. 
In fact, $A_7$ is completely determined by eqs.~\eqref{Gt} and~\eqref{g40_g41}:
\[
\label{A7_sech}
A_7 = 4w\cdot \mbox{sech}^2\left(\frac{t}{\tau}+u\right) - 2.
\]
Now, having defined $A_7$ in~\eqref{A7_sech}, there is $A_4$ 
in~\eqref{new_ksi} to be determined. 
According to the explicit form of $A_4$ in~\eqref{Theta}, 
the only yet unknown function there is $f_{41}(t)$. 
Recall that the main requirement is that $\pi$ becomes a conventional 
scalar field and General Relativity is restored in both distant past 
and future. Thus, eq.~\eqref{Theta} shows that in both asymptotic 
past and future $A_4 = -2 H$. 
We require that $(- A_4/2)$ is reasonably close to the Hubble 
parameter~\eqref{Hubble_choice} at all times. We achieve this by choosing
\[
\label{f_41}
f_{41}(t) = \frac{3w\cdot \mbox{sech}^2\left(\frac{t}{\tau}+u\right)}{2t\tau}\cdot \left[ t^2\cdot \tanh\left(\frac{t}{\tau} + u\right) +\tau^2\cdot\tanh\left(\frac{t}{\tau}\right)-t\cdot\tau \right].
\] 
This completes our definition of $F_4(\pi, X)$ in~\eqref{F4}.

\begin{figure}[h!]
\center{\includegraphics[width=0.65\linewidth]{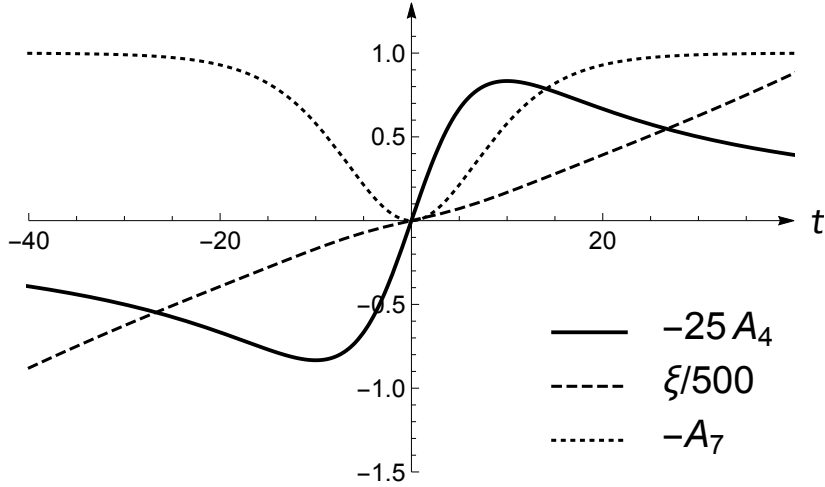}\hspace{1cm}}
\caption{A fine-tuned solution.}
\label{finetune}
\end{figure}

A comment on the fine-tuning issue is in order.
Under fine-tuning we mean the situation when the coefficient 
$A_4$ crosses zero at some moment of time $t_*$, while $A_7$ {\itshape touches} zero at $t=t_*$ and remains non-negative~\cite{Ijjas:2016tpn, Creminelli:2016zwa}. 
This situation is shown in Fig.~\ref{finetune}. In the case of fine-tuning eq.~\eqref{new_ksi} is satisfied, and both $\mathcal{C} \neq \infty$ 
and $\mathcal{A} \neq \infty$. 
In contrast, we aim at avoiding fine-tuning by introducing 
the constants $u$ and $w$ in eq.~\eqref{f_40}. 
We choose these constants in such a way that $A_4$ and $A_7$ 
cross zero at different times, see the bottom panel of Fig.~\ref{nogo}.

The only function still to be found is $F(\pi, X)$. We make use of $(00)$- and 
$(ij)$-components of equations of motion (see Appendix B) to
relate $f_0(t)$, $f_1(t)$ and $f_2(t)$:
\begin{subequations}
\label{backgr_eq}
\begin{align}
&f_0(t)-f_1(t)-3 f_2(t) + \frac{t}{3\tau\cdot (\tau^2+t^2)^2} \left\lbrace t\cdot\tau \phantom{\frac{t}{\tau}} \right.\\ \nonumber 
&\left.+ 6w\cdot \mbox{sech}^2\left(\frac{t}{\tau}+u\right) \left[\tau^2\cdot\tanh\left(\frac{t}{\tau}\right)-(2\tau^2+t^2) \tanh\left(\frac{t}{\tau}+u\right)\right]\right\rbrace = 0,\\
&3 f_0(t)+3 f_1(t)+3 f_2(t) + \frac{2 \tau^2-t^2}{(\tau^2+t^2)^2} = 0,
\end{align}
\end{subequations}
where we make use of eqs.~\eqref{g40_g41}, ~\eqref{f_40} and ~\eqref{f_41}.
From eqs.~\eqref{backgr_eq} one expresses $f_0(t)$ and $f_1(t)$ in terms of $f_2(t)$:
\begin{subequations}
\label{f_0andf_1}
\begin{align}
\label{explicit_f0}
& f_0(t) = \frac{1}{3 \tau \left(t^2+\tau^2\right)^2}\cdot \left[-\tau^3+3 \tau \left(t^2+\tau^2\right)^2\cdot f_2(t)-3 t \tau^2 w \cdot\mbox{sech}^2\left(\frac{t}{\tau}+u\right) \tanh\left(\frac{t}{\tau}\right)\right.\\\nonumber
&\left.+3 t^3 w\cdot \mbox{sech}^2\left(\frac{t}{\tau}+u\right) \tanh\left(\frac{t}{\tau}+u\right)+6 t \tau^2 w\cdot \mbox{sech}^2\left(\frac{t}{\tau}+u\right) \tanh\left(\frac{t}{\tau}+u\right)\right],\\
\label{explicit_f1}
& f_1(t) = 
-\frac{1}{3 \tau \left(t^2+\tau^2\right)^2}\cdot \left[\tau^3+6 \tau (t^2+\tau^2)^2 \cdot f_2(t)-3 t \tau^2 w \cdot \mbox{sech}^2\left(\frac{t}{\tau}+u\right) \tanh\left(\frac{t}{\tau}\right)\right.\\\nonumber
&\left.+3 t^3 w\cdot \mbox{sech}^2\left(\frac{t}{\tau}+u\right) \tanh\left(\frac{t}{\tau}+u\right)+6 t \tau^2 w\cdot \mbox{sech}^2\left(\frac{t}{\tau}+u\right) \tanh\left(\frac{t}{\tau}+u\right) -t^2 \tau \right].
\end{align}
\end{subequations}
The final step is to choose $f_2(t)$ in such a way that $\mathcal{A} > 0$ (no ghosts in the scalar sector of~\eqref{action_no-go}) and $\mathcal{A} > \mathcal{C}$ (scalar perturbations propagate at 
subluminal speed). The only unconstrained coefficient left in the definition of $\mathcal{A}$ in~\eqref{mathcal_A} is $A_3$, which, according to~\eqref{Sigma}, involves the yet 
undetermined function $f_2(t)$. 
To satisfy both $\mathcal{A} > 0$ and $\mathcal{A} > \mathcal{C}$ we choose
$f_2(t)$ in such a way that the term involving $A_3$ is always dominating in~\eqref{mathcal_A}.
Since we chose $A_5=2$ and have $A_1=-\frac32 A_5 = -3$
(see eq.\eqref{A_5}), it is sufficient to choose $A_3$ as follows:
\[
\label{Sigma_approx}
A_3 = \left[ 1 + \left(\frac{t}{\tau}\right)^2\right]^{-2},
\] 
which gives
\[
\begin{aligned}
\label{f_2}
f_2(t) & =\frac{1}{12 \tau (t^2+\tau^2)^2}\cdot\left\lbrace\tau^3+3 \tau^5+4 t w\cdot \mbox{sech}^2\left(\frac{t}{\tau}+u\right) \left[-4 t \tau +9 \tau^2 \tanh\left(\frac{t}{\tau}\right)\right.\right. \\\nonumber
&\left.\left.+3 (t^2-2 \tau^2) \tanh\left(\frac{t}{\tau}+u\right)\right]\right\rbrace.
\end{aligned}
\] 
This completes  the reconstruction of the Lagrangian functions~\eqref{lagr_func_choice}.

\end{document}